\documentclass[a4paper]{article}

\usepackage{makeidx}         
\usepackage{graphicx}        
\usepackage{multicol}        
\usepackage[bottom]{footmisc}

\usepackage{cite}

\usepackage{color}

\DeclareTextSymbol{\degre}{OT1}{23}

\title{Predictive use of the Maximum Entropy Production principle for Past and Present Climates}
\author{C. Herbert, D. Paillard}
\date{}

\begin{document}

\maketitle
\footnotetext{C. Herbert and D. Paillard,  Laboratoire des Sciences du Climat et de l'Environnement, IPSL, CEA-CNRS-UVSQ, UMR 8212, 91191 Gif-sur-Yvette, France, corentin.herbert@lsce.ipsl.fr.}

\abstract{In this chapter, we show how the MEP\index{Maximum Entropy Production} hypothesis may be used to build simple climate models\index{Climate Models} without representing explicitly the energy transport by the atmosphere. The purpose is twofold. First, we assess the performance of the MEP hypothesis by comparing a simple model with minimal input data to a complex, state-of-the-art General Circulation Model\index{Climate Models!General Circulation Models}. Next, we show how to improve the realism of MEP climate models\index{Climate Models!MEP models} by including climate feedbacks\index{Climate Feedbacks}, focusing on the case of the water-vapour feedback\index{Climate Feedbacks!Water-vapour Feedback}. We also discuss the dependence of the entropy production\index{Entropy production} rate and predicted surface temperature on the resolution of the model.}

\section{Introduction}

Although it is not straightforward to define what climate is precisely, one may suggest that what we call \emph{the climate system} is made up of the atmosphere, the oceans, the cryosphere, the biosphere and the lithosphere \cite{PeixotoPOC}. The different components interact in various ways, and their relative importance depends on the question asked. For instance in numerical weather prediction\index{Numerical Weather Prediction}, taking place on a timescale of a few days, the main dynamical component is the atmosphere and all the other components may be regarded as prescribed. On the contrary, the evolution of climate on very long timescales (of the order of tenths or hundreds million year) is essentially determined by the exchanges of carbon between the land, the oceans and the atmosphere.

The distribution of surface temperature is of primary interest. It depends on a large number of factors, such as the composition of the atmosphere (upon which the radiative energy exchanges depend), the circulation of the atmosphere and oceans, the ocean salinity, the presence of ice-sheets, the type of terrestrial vegetation cover,... State-of-the-art climate models, usually referred to as \emph{General Circulation Models} (GCMs)\index{Climate Models!General Circulation Models}, now include many of the above factors (the term \emph{Earth System Models}\index{Climate Models!Earth System Models} is starting to emerge). 

However, not all this complexity is necessary to obtain a rough estimate of the temperature of a planetary atmosphere: perhaps the simplest approach is to balance the incoming solar radiation with the outgoing planetary radiation. Again this can be done at various levels of accuracy, depending on the knowledge we have of the concentration of the radiatively active constituents of the atmosphere (e.g. water-vapour and carbon dioxide). Imposing a local radiative equilibrium is in fact misleading: latitudinal and vertical differential heating trigger atmospheric motions, which carry heat to mitigate the temperature gradients that would exist at radiative equilibrium. The resulting energy transport term can be parametrized (for instance as a diffusion process with empirical diffusivity) as a function of the temperature distribution, so that we can solve the model without resolving explicitly the motions of the atmosphere. Such models, consisting of a radiative model and a parameterization of the energy transport by the atmosphere are called \emph{Energy Balance Models} (EBMs)\index{Climate Models!Energy Balance Models}. Alternatively, one may solve the fluid dynamics problem and compute explicitly the velocity field: this is what GCMs do. The hierarchy of climate models, ranging from simple EBMs to complex GCMs, also comprises the so-called \emph{intermediate complexity models} (EMIC)\index{Climate Models!Earth Models of Intermediate Complexity}, which offer a variety of simplified representations of the atmospheric and oceanic circulation and other phenomena \cite{McGuffieHendersonSellersBook}. The main interest of EMICs is their relatively low computational cost, compared to GCMs\index{Climate Models!General Circulation Models}, which make them particularly suitable for the study of palaeoclimates\index{Palaeoclimate}. Indeed, the timescales involved in such problems reduce the role of GCMs to simulating snapshots. Both GCMs and EMICs require a certain amount of \emph{parameter tuning}. This is sometimes a problem when studying past climates for which little data is available on which to base adjustment procedures, and even more so for other planetary climates, where many features differ tremendously from the terrestrial conditions on which the empirical parameterizations were tested.

Nevertheless, the laws of physics remain the same when going back into time or out into the cosmos. The three branches of physics which play a fundamental part in setting the climate of a planet are radiation physics \cite{GoodyBook}, fluid dynamics \cite{HoltonBook,PedloskyGFD} and thermodynamics \cite{AmbaumBook}. One fundamental principle which is always present, even in simple models like EBMs, is the first law of thermodynamics\index{First Law of Thermodynamics}, because it describes the exchanges of energy in a system. To energy exchanges are associated equilibrium temperature distributions. On the other hand, even in the most sophisticated climate models to date, the second law of thermodynamics\index{Second Law of Thermodynamics}, which also describes the exchanges of energy in a system but in a qualitative rather than quantitative way, is not taken into account.
When subgrid-scale parameterizations\index{Sub-grid Parameterizations} are involved, classical models may even violate the second law of thermodynamics \cite{Holloway2004}. It has also been suggested that spurious sources of entropy production\index{Entropy production}  could lead to a global cold bias in climate models \cite{Johnson1997}. Henceforth, a number of diagnostic tools emerged to study the thermodynamic properties of climate models \cite{Lucarini2009b,Boschi2012}. 
Besides, postulating that the system chooses the steady-state with maximum entropy production given certain constraints leads to a variational problem which has proved very efficient for predictive use. This is the so-called \emph{Maximum Entropy Production principle} \cite{kleidonlorenzbook,Ozawa2003,Martyushev2006}.
We shall not discuss here the theoretical foundations (or lack thereof) of this hypothesis (see \cite{Dewar2003,Dewar2004,Grinstein2007,Bruers2007}), but only its consequences for climate modelling. Hitherto, mainly two approaches have been developed.
One point of view is that the MEP principle\index{Maximum Entropy Production} can be useful to select the value of adjustable parameters in empirical parameterizations from existing models, in an \emph{objective} way \cite{Kleidon2003,Ito2004,Kleidon2006,Kunz2008,Pascale2012}\index{Sub-grid Parameterizations}. In the second approach, the purpose is to build simple climate models based on the MEP hypothesis for describing unresolved processes. We shall present the latter approach in this chapter. After briefly reviewing earlier attempts (Sect. \ref{paltridgesection}) we build a MEP climate model devoid of \emph{ad hoc} assumptions and we show how to include feedbacks like the water-vapour feedback (Sect. \ref{nefmodelsection}). The model is then tested for pre-industrial and Last Glacial Maximum\index{Palaeoclimate!Last Glacial Maximum} conditions (Sect. \ref{resultssection}).

\section{The Paltridge model}\label{paltridgesection}

A typical one-dimensional EBM\index{Climate Models!Energy Balance Models} consists of a certain number of \emph{boxes}, representing latitudinal zones, characterized by a single temperature. Each box receives energy from the outside in the form of solar radiation, and radiates back to space in the longwave domain. The difference of these two terms, which is usually called the \emph{radiative budget} of the box, does not necessarily vanish: there are also energy exchanges with the neighbouring boxes due to atmospheric (and oceanic) transport of heat. Hence, for box $i$, the total energy budget reads

\begin{equation}
{c_p}_i \frac{d T_i}{dt} = R_i + \gamma_i,
\end{equation}
where ${c_p}_i,T_i,R_i$ and $\gamma_i$ denotes respectively the heat capacity, temperature, radiative budget and atmospheric (or oceanic) convergence for box $i$. A \emph{radiative scheme} provides $R_i$ as a function of $T_i$: e.g. $R_i=\xi_i S - \epsilon_i \sigma T_i^4$ where $S$ is the solar constant, $\xi_i$ represents the projection of the surface of the latitude belt onto the sphere centered on the sun, $\sigma$ is the Stefan-Boltzmann constant and $\epsilon_i$ the emissivity of the surface. In such a radiative scheme, the greenhouse effect is not taken into account. In contrast, there is no simple expression for $\gamma_i$ which can be justified from first principles. A standard \emph{parameterization} in this context is to assume a diffusion-like term, but there is no justification for this hypothesis and the diffusion coefficient has to be chosen empirically.\index{Sub-grid Parameterizations}

Paltridge \cite{Paltridge1975} suggested a model\index{Climate Models!MEP models}, with a more elaborate radiative scheme - involving in particular a cloud cover variable $\theta_i$ in each box - than our above example, in which $\gamma_i$ is not empirically parameterized as a function of the temperatures $T_i$, but instead satisfies a maximum entropy production principle. He postulates that the steady-state temperature distribution $T_i$ is such that the material entropy production rate\index{Entropy production}  $\sigma=\sum_i \frac{\gamma_i}{T_i}$ is maximum, subject to the global steady-state constraint $\sum_i \gamma_i = 0$.  At steady state, $\gamma_i=-R_i$ and $\sigma$ is a function of the temperatures $T_i$ only. At steady-state, the distributions of temperature, cloud cover, atmospheric and oceanic meridional fluxes obtained are in striking accordance with observations. In spite of this apparent success, some major criticism remain. First of all, the planetary rotation rate is believed to be a major driver of the latitudinal distribution of temperatures, but it does not appear at all in Paltridge's model. Besides, it is clear that the principle does not hold in the case of a planet without atmosphere (see \cite{Fukumura2012}). One may thus wonder if it is not pure coincidence that it seems to apply to the Earth's atmosphere \cite{Rodgers1976}. Last but not least, there is no theoretical justification for the principle of maximum entropy production.

The thread was taken up in a series of papers \cite{Paltridge1978,Grassl1981,Wyant1988,Gerard1990}, verifying Paltridge's results in different variants of the original model, but the fundamental objections mentioned above remained unanswered. More recently, Lorenz \cite{Lorenz2001} added some support to the idea that the agreement between the model and observations is not a coincidence, by showing that it gives acceptable results for Titan and Mars as well. The question of the independence with respect to the planetary rotation rate was also adressed by Jupp \cite{Jupp2010} in a MEP model with a simple parameterization of atmospherics dynamics.
Nevertheless, one fundamental concern remains: the Paltridge model and its variants still contain a large number of parameterizations, \emph{ad hoc} hypothesis and empirical coefficients, for instance in the radiative scheme, in the cloud parameterization or in the treatment of surface heat fluxes (maximum convective hypothesis). Is it possible to get rid of these potential biases to assess the intrinsic value of the MEP conjecture\index{Maximum Entropy Production} in the climate modelling framework ? This is the question we address in the next section.

\section{A simple MEP model with water-vapour feedback}\label{nefmodelsection}

\subsection{NEF Radiative scheme}

A possible strategy to assess the degree of coincidence in Paltridge's results may be to build a MEP model\index{Climate Models!MEP models} devoid of any \emph{ad-hoc} parameter and assumptions. To that end, we suggest a new radiative scheme based on the Net Exchange Formulation (NEF)\index{Net Exchange Formulation}, which only involves physical quantities (values of which are known \emph{a priori}). Following \cite{Herbert2011b}, we introduce a two dimensional model with two layers: for each grid point characterized by a latitude and a longitude, there is a surface layer with a temperature $T_g$ and an atmospheric layer with a temperature $T_a$. 

\begin{figure}
\includegraphics[width=\textwidth]{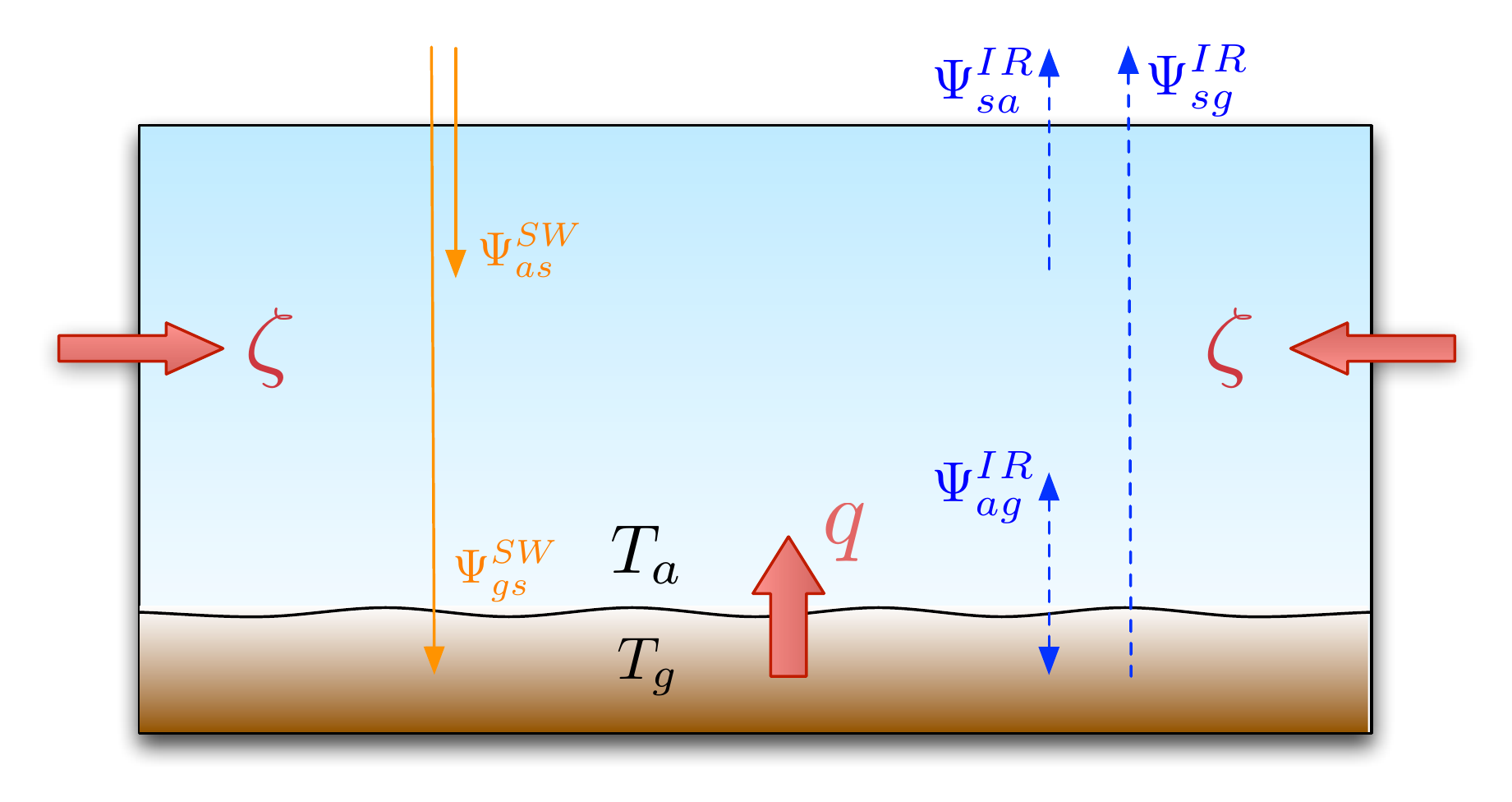}
\caption{One grid cell of a two-layer MEP model. The surface layer has temperature $T_g$ and exchanges heat $q$ (thick solid red arrow) with the overlying atmospheric layer of temperature $T_a$. Both layers absorb solar radiation (thin solid yellow arrows) and emit and absorb longwave radiation (thin dashed blue arrows). The atmospheric layer exchanges energy with the surrounding cells: the convergence of the atmospheric heat flux is $\zeta$ (thick solid red arrow).}
\end{figure}

Each layer absorbs an amount of solar radiation ($\Psi_{gs}^{SW}$ for the surface layer and $\Psi_{as}^{SW}$ for the atmosphere) given by:
\begin{eqnarray}
\Psi_{gs}^{SW}&=&(\bar{s}(\alpha_g)-s)(1-\alpha_g)\xi S, \label{psisweqn1}\\
\Psi_{as}^{SW}&=&(s+\alpha_gs^*)\xi S,\label{psisweqn}
\end{eqnarray}
where $S$ is the solar constant, $\xi$ the projection of the cell area onto the sphere, $\alpha_g$ the surface albedo, and the coefficients $s, s^*$ and $\bar{s}$ are adapted from the classical Lacis and Hansen scheme \cite{Lacis1974}:
\begin{eqnarray}
\bar{s}(\alpha_g)&=&0.353+\frac{0.647-\bar{R_r}(\xi)-A_{oz}(Mu_{O_3})}{1-\bar{\bar{R_r^*}}\alpha_g },\\
s&=&A_{wv}(M\tilde{u}),\\
s^*&=&A_{wv}\left(\left(M+\frac{5}{3}\right)\tilde{u}\right)-A_{wv}(M\tilde{u}).
\end{eqnarray}
Here $u_{O_3},\tilde{u}$ represent respectively the vertically integrated ozone and water vapour density (including pressure scaling \cite{Stephens1984}), $M$ accounts for the slant path of solar rays, $\bar{R_r}(\xi)$ and $\bar{\bar{R_r^*}}$ account for Rayleigh scattering in the atmosphere, and $A_{oz},A_{wv}$ are absorption functions for ozone and water vapour. See \cite{Lacis1974,Stephens1984,Herbert2011b} for details.

The long-wave radiative exchanges can be written in a simple form using the Net Exchange Formulation \cite{Dufresne2005}. The surface layer and the atmosphere exchange a net amount of energy $\Psi_{ag}^{IR}$ through infrared radiation, while the surface and the atmosphere radiate respectively $\Psi_{sg}^{IR}$ and $\Psi_{sa}^{IR}$ to space (see \cite{Herbert2011b} for a derivation):
\begin{eqnarray}\label{psilweqn}
\Psi_{ag}^{IR}&=&t (T_g) \sigma T_g^4-t(T_a)\sigma T_a^4,\\
\Psi_{sa}^{IR}&=&t (T_a) \sigma T_a^4,\\
\Psi_{sg}^{IR}&=&\left(1-\frac{t (T_g)}{\mu}\right)\sigma T_g^4,
\end{eqnarray}
where $\mu$ is the Elsasser factor arising from the angular integration, and $t(T)=\mu\left(1-\int_0^{+\infty} \frac{B_\nu(T)}{\sigma T^4} \tau_\nu d\nu\right)$ represents the emissivity of the atmosphere ($B_\nu$ is the Planck function). The transmission function $\tau_\nu$ depends on the vertical profiles of absorbing gases, pressure and temperature: $\tau_\nu=\exp \left( - \frac{1}{\mu} \int_0^H k_\nu (z)dz\right)$, where $k_\nu$ is the absorption coefficient, and $H$ the total height of the atmosphere. 
To sum up, the only parameters required by the radiative scheme are the vertically integrated concentrations of water vapour $\tilde{u}$, carbon dioxide $u_{CO_2}$ (they determine $k_\nu$), ozone $u_{O_3}$ and the surface albedo $\alpha_g$.

The steady-state condition for each box reads, for every grid point:
\begin{eqnarray}\label{energybalanceeq1}
\Psi_{gs}^{SW}+\Psi_{as}^{SW}-\Psi_{sg}^{IR}-\Psi_{sa}^{IR}+\zeta&=&0,\\
\Psi_{gs}^{SW}-\Psi_{ag}^{IR}-\Psi_{sg}^{IR}-q&=&0,\label{energybalanceeq2}
\end{eqnarray}
where $\zeta$ is the horizontal convergence of atmospheric heat fluxes and $q$ the surface to atmosphere heat flux. The total material entropy production\index{Entropy production}  is given by
\begin{equation}
\sigma_M(\{T_{a,ij},T_{g,ij}\})=\sum_{i=1}^{N_{lat}}\sum_{j=1}^{N_{lon}} \left(\frac{q_{ij}}{T_{a,ij}}-\frac{q_{ij}}{T_{g,ij}}+\frac{\zeta_{ij}}{T_{a,ij}}\right)A_{ij},
\label{sigmadefeq}
\end{equation}
where $A_{ij}$ is the area of the grid cell in position $(i,j)$ and $q_{ij},\zeta_{ij}$ are functions of $T_{a,ij},T_{g,ij}$ given by (\ref{energybalanceeq1})-(\ref{energybalanceeq2}) . We are interested in the fields that maximize $\sigma_M$ while satisfying the global constraint $\sum_{i,j} A_{ij} \zeta_{ij} =0$, which can be translated into an unconstrained variational principle using Lagrange multipliers.

\subsection{Different versions of the model}\label{versionssection}

The MEP model\index{Climate Models!MEP models} described in the previous section requires only physical parameters as an input. In a first step, we compute the horizontal distribution of $\tilde{u}$ (vertically integrated water vapour density) and $u_{O_3}$ by linear interpolation of standard atmospheric profiles \cite{McClatchey1972} (depending only on the latitude). To compare with the results of Paltridge, we also assume that the coefficients $t(T)$ in Eqs. \ref{psilweqn} are fixed, with a prescribed reference temperature $T_{ref}$ (dependent on the latitude) also computed from the standard profiles \cite{McClatchey1972} (version v0 in table \ref{versiontable}). 
However, the assumption of constant $t(T)$ coefficient is  very unrealistic: the shift in the Planck spectrum associated with a variation in temperature of the surface or atmospheric layer has a strong impact on the optical properties of the atmosphere. In version v1, we retain the dependence of the emissivity of the atmosphere on surface and atmospheric temperatures. 
Besides, fixing the profiles of water-vapour and ozone is also a restrictive hypothesis, especially in view of potential applications to different climates for which standard profiles are not well known. As far as ozone is concerned, we can simply examine a version of the model in which we completely ignore ozone (version v2).
For water-vapour, the situation is slightly more complicated: the atmospheric temperature is linked via the Clausius-Clapeyron relation to the water vapour content, which itself feeds back onto the temperature via the greenhouse effect. Yet, in the previous versions (v0-v2) of the MEP model, we kept fixed the absolute amount of water vapour in the atmosphere, independently of the temperature. In version v3, we fix the relative humidity\index{Relative Humidity} $RH=P_{H_2O}/P_{sat}(T)$. The vertically integrated density of water vapour is related to the relative humidity, temperature and pressure profiles through:
\begin{equation}\label{uh2orheq}
u_{H_2O}^* = \frac{1}{g} \frac{M_{H_2O}}{M_{air}} \int_0^{P_s} RH \times P_{sat}(T) \frac{dp}{p},
\end{equation}
where $M_{H_2O},M_{air}$ are the molar masses of water and air, $g$ is the gravity and $P_s$ the surface pressure.
In our model with one atmospheric layer, we may assume that the relative humidity is uniformly distributed in each atmospheric cell, with a vertical extent equal to the scale height for water vapour. Relation (\ref{uh2orheq}) then becomes $u_{H_2 O}^* \approx M_{H_2O}/(g M_{air}) \times RH \times P_{sat}(T)$ (version 3). The different versions are summarized in Table \ref{versiontable}. The purpose of comparing these different versions of the model is at the same time to test the impact of reducing the quantity of input parameters (no $T_{ref}$, no $u_{O_3}$) and to improve the realism (Planck spectrum, water-vapour feedback)\index{Climate Feedbacks!Water-vapour Feedback}.

\begin{table}
\centering
\begin{tabular}{|c|cccc||cc|}
\hline
Model Version & $\tilde{u}$ & $u_{O_3}$ & $u_{CO_2}$ & $t(T)$ & $\langle T_{PI} \rangle$ (\degre C) & $\langle T_{LGM}-T_{PI}\rangle$ \\
\hline
MEP v0 & MC & MC & 280ppmv & $T=T_{ref}$ (MC) & 22.9 & -1.98\\
MEP v1 & MC & MC & 280ppmv & $T=T_a,T_g$ & 22.3 & -1.84\\
MEP v2 & MC & 0 & 280ppmv & $T=T_a,T_g$ & 22.5 & -1.84\\
MEP v3 & $u^*(T_a)$ & 0 & 280ppmv & $T=T_a,T_g$ & 19.9 & -2.9\\
\hline
IPSL & - & - & 280 ppmv & - & 15.7 & -2.53\\
\hline
\end{tabular}
\caption{Different versions of the MEP model and the resulting global mean surface temperature for pre-industrial (PI) and last glacial maximum (LGM) climates, compared to GCM\index{Climate Models!General Circulation Models} runs with the IPSL\_CM4 model. "MC" stands for the integrated standard McClatchey profiles, and the angular brackets mean global average. See Sect. \ref{versionssection} for the definition of the different versions and Sect. \ref{resultssection} for the discussion of the results. }\label{versiontable}
\end{table}

\subsection{Water-vapour feedback and multiple steady states}\index{Climate Feedbacks!Water-vapour Feedback}

The physical quantities involved in the climate system are related in many ways, so that a change in one of these quantities can have an influence on another one, feeding back onto the original quantity, either moving it closer (negative feedback) or farther (positive feedback) from its initial value. A classical example of positive feedback is the water-vapour feedback. If the temperature increases locally, the water vapour saturation pressure will increase so that more water (if available) may evaporate in the atmosphere, leading to stronger greenhouse effect and thus further increase of the temperature. Feedbacks of this sort can lead to multiple equilibria, bifurcations and hysteresis phenomena. For a given relative humidity\index{Relative Humidity} distribution, equilibrium states with radically different temperatures are simultaneously possible \cite{Renno1997}. The water-vapour feedback has been shown to play a major part in important climate problems \cite{Pierrehumbert2002}, exactly like feedbacks of different natures \cite{Roe2007,Lenton2008}.\index{Climate Feedbacks}
Hence, it is essential to be able to represent them correctly in a climate model. 
In the context of MEP models\index{Climate Models!MEP models}, it was shown in \cite{Herbert2011a} that the ice-albedo feedback\index{Climate Feedbacks!Ice-albedo Feedback} gives rise to multiple local maxima in the entropy production rate\index{Entropy production} , corresponding to the multiple equilibria\index{Multiple Equilibria} that appear in a traditional EBM\index{Climate Models!Energy Balance Models} (see also \cite{Boschi2012}). Here, we observe multiple local maxima of the entropy production rate in a certain range of solar constant and relative humidity.
One great advantage of MEP\index{Maximum Entropy Production} is the small computational cost of maximizing a function as compared to integrating a complex differential equation. Of course this is no longer true if the function, or the submanifold on which to search for the maximum, becomes too complicated.
Already, in the presence of multiple maxima, this difficulty has to be dealt with as the steady-state selected by the maximization algorithm may depend on the initial value. To avoid being trapped in an irrelevant state, several methods may be investigated. 
First it is possible to further restrict the manifold defined by the constraints to ensure that it contains only one local maximum of the entropy production. In the case of the water-vapour feedback in our two-layer model, solving the radiative balance for the whole column in terms of the atmospheric temperature may lead to several solutions. Selecting systematically one of them before computing the entropy forces the system to remain on the portion of interest in phase space. This is the technique that we use here.
Alternatively, introducing the time dimension and assuming that at each time step, the system maximizes instantaneous entropy production\index{Entropy production}  with an additional term corresponding to time derivatives, it was suggested in \cite{Herbert2011a} to use \emph{relaxation equations} as a numerical algorithm to compute the final state (see also \cite{Vallino2012}).

\section{Results: Present and Last Glacial Maximum climates}\label{resultssection}

\begin{figure}
\includegraphics[width=\textwidth]{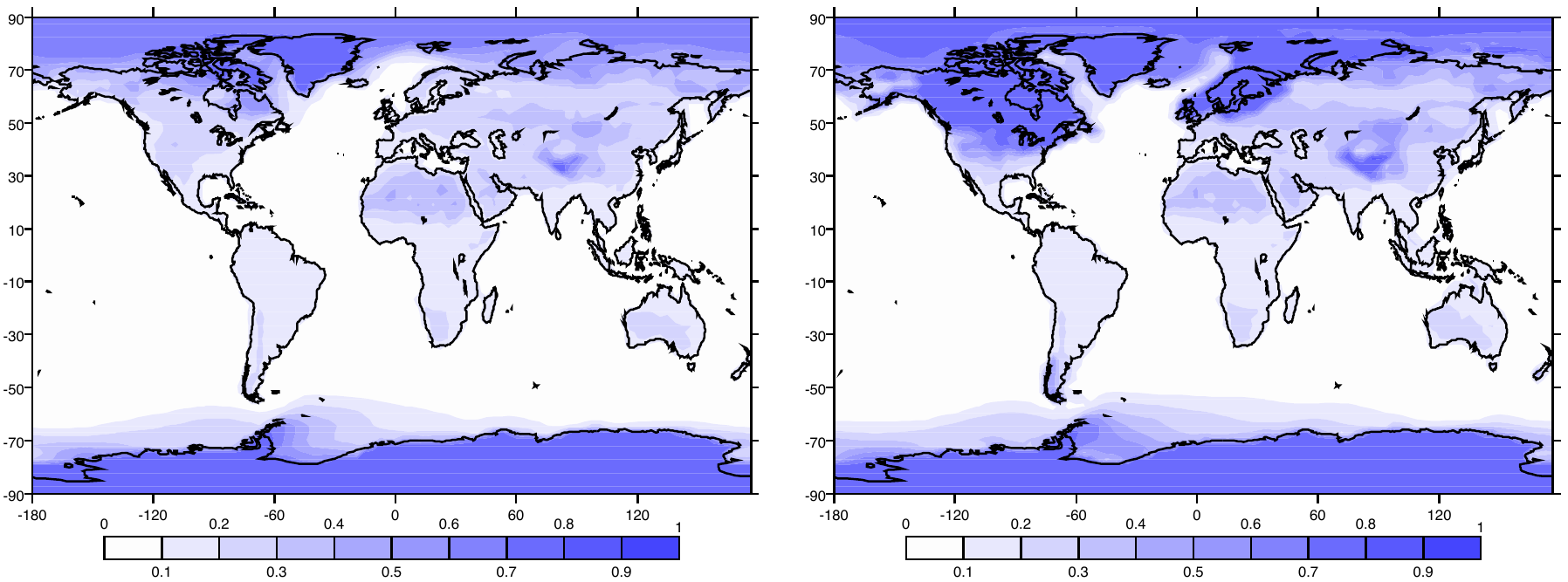}
\caption{Surface albedo $\alpha_g$ in the IPSL model, for pre-industrial (left) and Last Glacial Maximum (right) conditions.}\label{albfig}
\end{figure}

We compared the surface temperature distribution obtained from MEP with that obtained from a state-of-the-art GCM\index{Climate Models!General Circulation Models}, the IPSL\_CM4 model. The IPSL model is a coupled atmosphere-ocean model \cite{Marti2010} used for the Fourth Assessment Report (AR4) of the Intergovernmental Panel on Climate Change (IPCC) \cite{ipccAR4}. For pre-industrial climate, the forcings in the IPSL model are: pre-industrial greenhouse gas concentration (CO$_2$=280 ppm, CH$_4$=760 ppb, N$_2$O=270 ppb), insolation, coastlines, topography and land-ice extent. The surface albedo is computed from the IPSL\_CM4 pre-industrial simulation and used as a forcing for the MEP model (Fig. \ref{albfig}, left).

\begin{figure}
\includegraphics[width=\textwidth]{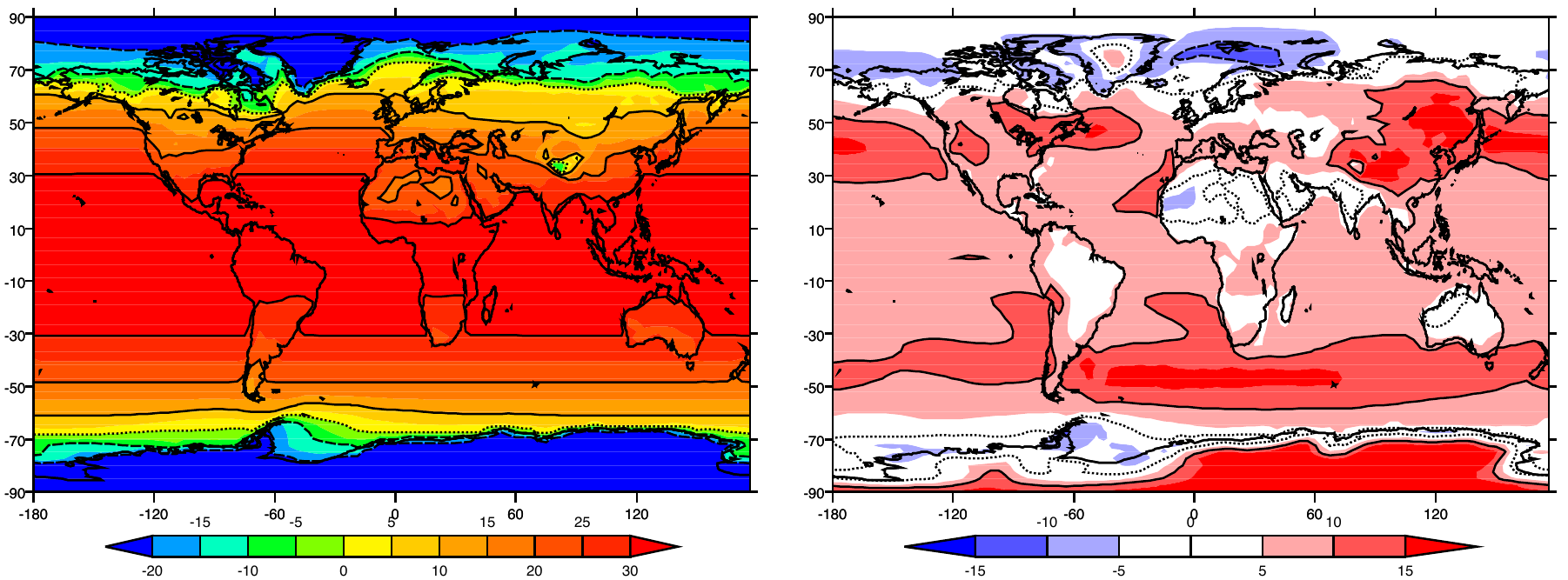}
\caption{Left: surface temperature $T_g$ for pre-industrial conditions obtained with the MEP model (version v0). Right: Difference between the surface temperature $T_g$ in the MEP model and the IPSL model for pre-industrial conditions.  Contour lines interval is 10\degre C, positive contours are drawn in solid lines, negative contours in dashed lines and the null contour as a dotted line.}\label{PIsurftempfig}
\end{figure}

\begin{figure}
\begin{center}
\includegraphics[width=0.5\textwidth]{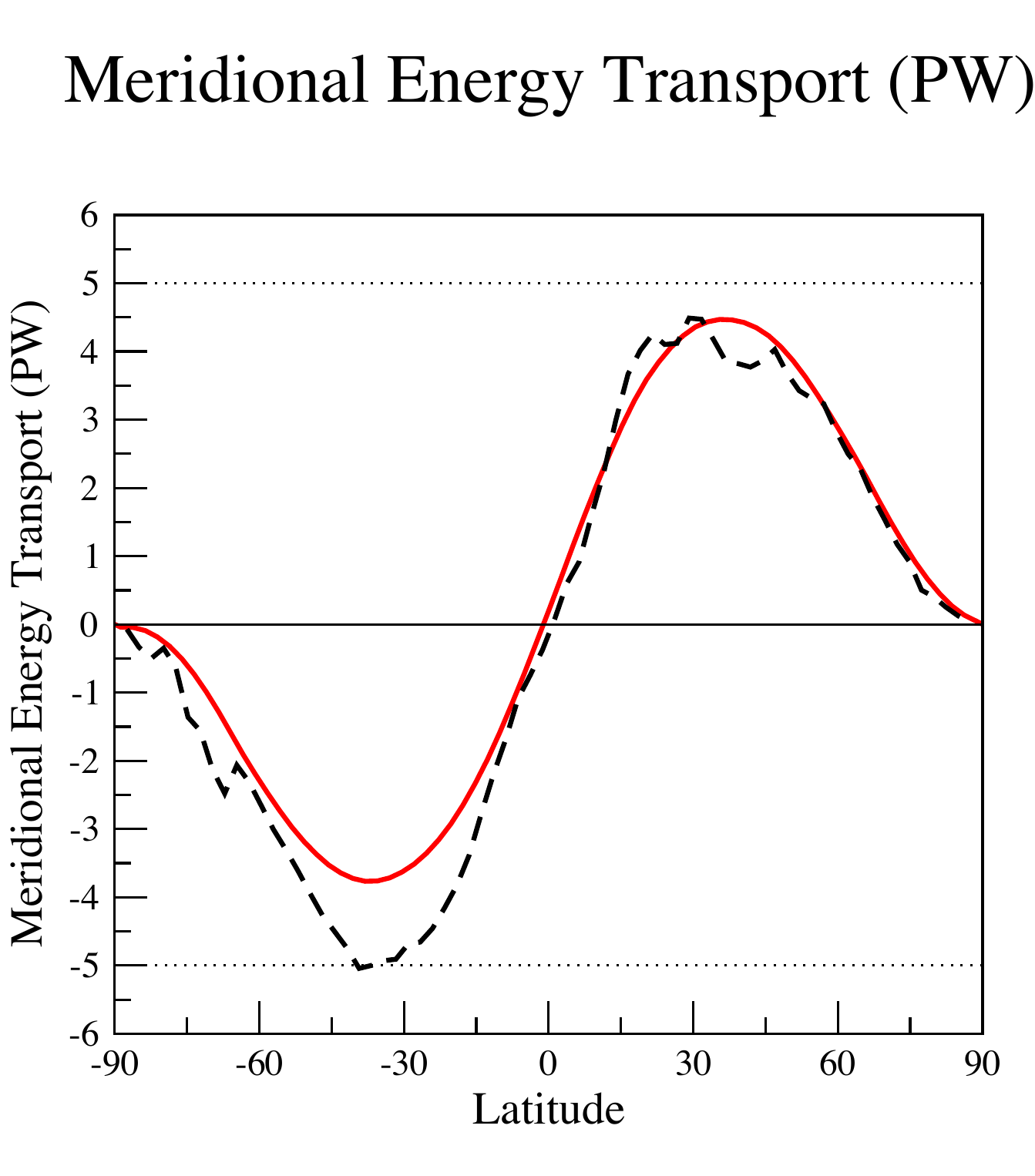}
\caption{Meridional energy transport as a function of the latitude in the MEP model (version v0, solid red line) and the IPSL model (dashed black line), for pre-industrial conditions.}\label{PItransportfig}
\end{center}
\end{figure}

The surface temperature distribution obtained with the MEP model is represented in Fig. \ref{PIsurftempfig} along with the difference between the MEP model and the IPSL model. The global mean surface temperature for the MEP model is $\langle T_{PI} \rangle = 22.9$\degre C. By comparison, $\langle T_{PI}\rangle $ in the IPSL simulation is approximately $7$\degre C lower (Table \ref{versiontable}); as Fig. \ref{PIsurftempfig} reveals, the major part of this difference comes from areas where the cloud cover is important, or elevated areas like the Antarctica. It is shown in \cite{Herbert2011b} that a crude estimation of the effect of clouds and elevation suffices to explain the major part of the difference with the IPSL model.
Figure \ref{PItransportfig} shows the meridional energy transport as a function of latitude for both the MEP model and the IPSL model for pre-industrial conditions. The agreement is remarkable given the simplicity of the MEP model.

\begin{figure}
\includegraphics[width=\textwidth]{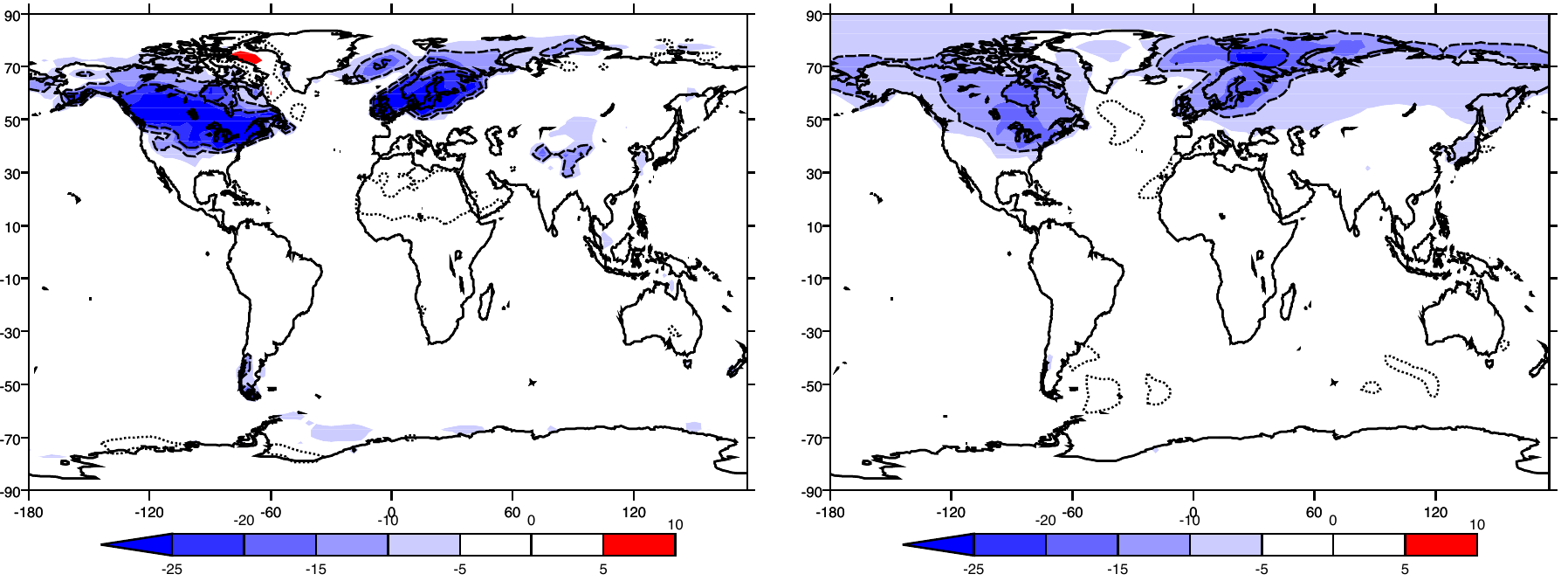}
\caption{Surface temperature difference between the Last Glacial Maximum and the Pre-Industrial, in the MEP model (left, version v0) and in the IPSL model (right). Contour lines space is 10\degre C, positive contours are drawn in solid lines, negative contours in dashed lines and the null contour as a dotted line.}\label{LGMsurftempfig}
\end{figure}

One advantage of the reformulation of the Paltridge model presented here is that due to the absence of \emph{ad-hoc} parameters, it is possible to test the model on climates other than the Pre-Industrial period. For instance, it is possible to change the surface albedo to take into account the variations of ice or vegetation extent. A time period which is largely documented and for which simulations with GCMs\index{Climate Models!General Circulation Models} are available is the \emph{Last Glacial Maximum} (LGM)\index{Palaeoclimate!Last Glacial Maximum}. It corresponds to the time during the last glacial period when the ice-sheets extent was maximum, roughly 21 000 years ago \cite{CrowleyBook}.
At that time, large ice-sheets covered North America and Northern Europe, and the global mean temperature was approximately $5$\degre C lower than present. In the MEP model, it is only possible to take into account the effect in surface albedo due to the presence of the ice-sheets at the LGM (Fig. \ref{albfig}, right), and not, for instance the associated topography effect. To ensure the comparison with the IPSL model is as direct as possible, we use a simulation where only the albedo effect is taken into account in the GCM. The resulting surface temperature difference between the LGM and the PI is shown in Fig. \ref{LGMsurftempfig} for both models. The global mean difference is $\approx -2$\degre C in the case of the MEP model and $\approx -2.5$\degre C for the IPSL model. However, in the IPSL model the temperature anomaly spreads over a large area in the Northern Hemisphere, while in the MEP model, it concentrates over the area where the ice-sheets are.

\begin{figure}
\begin{center}
\includegraphics[width=0.5\textwidth]{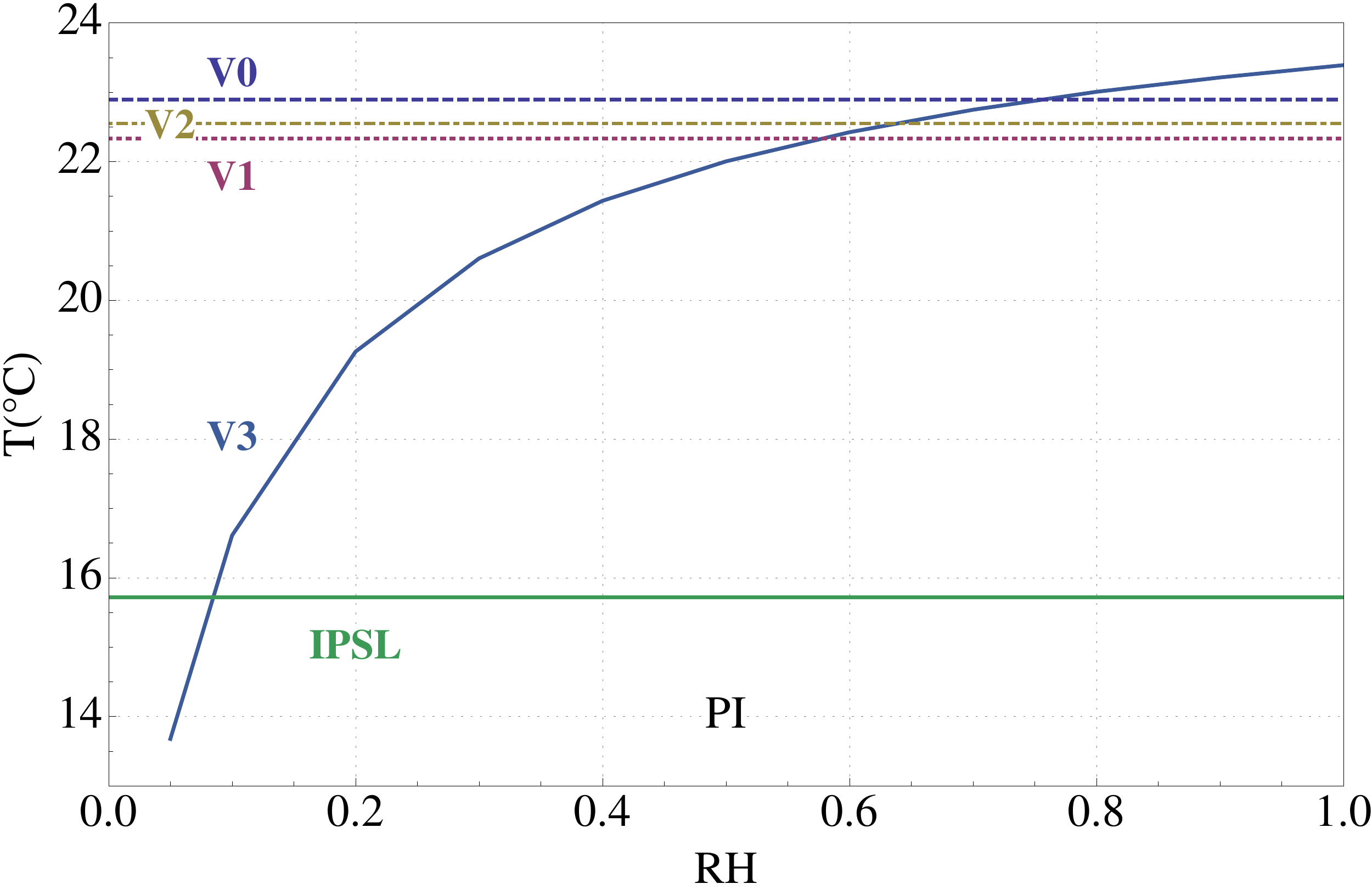}
\caption{Solid blue curve: Global mean surface temperature $T_g$ as a function of relative humidity (with a homogeneous distribution). The horizontal lines indicate the temperature obtained by fixing the absolute humidity in the MEP model, versions v0 (dashed blue), v1 (dotted red) and v2 (dashed-dotted yellow), and for the IPSL model (green solid line). }\label{TsurfRHfig}
\end{center}
\end{figure}

Table \ref{versiontable} compares the global mean surface temperatures obtained using the different models, for both Pre-Industrial and Last Glacial Maximum\index{Palaeoclimate!Last Glacial Maximum} conditions. Including the interactive Planck spectrum (version v1 compared to version v0) leads to a slight cooling (0.6\degre C) and a smaller albedo sensitivity, while turning off the ozone (version v2 compared to version v1) yields a very small warming (0.2\degre C) and does not change the sensitivity. 
Figure \ref{TsurfRHfig} shows the dependence of the global mean surface temperature on relative humidity\index{Relative Humidity}. For simplicity, a horizontally homogeneous relative humidity distribution is used. The global mean surface temperature spans a wide interval, between approximately 14\degre C and 24\degre C. In particular, it encompasses the global mean surface temperature obtained with other versions of the MEP model and with the IPSL model.

\begin{figure}
\includegraphics[width=0.5\textwidth]{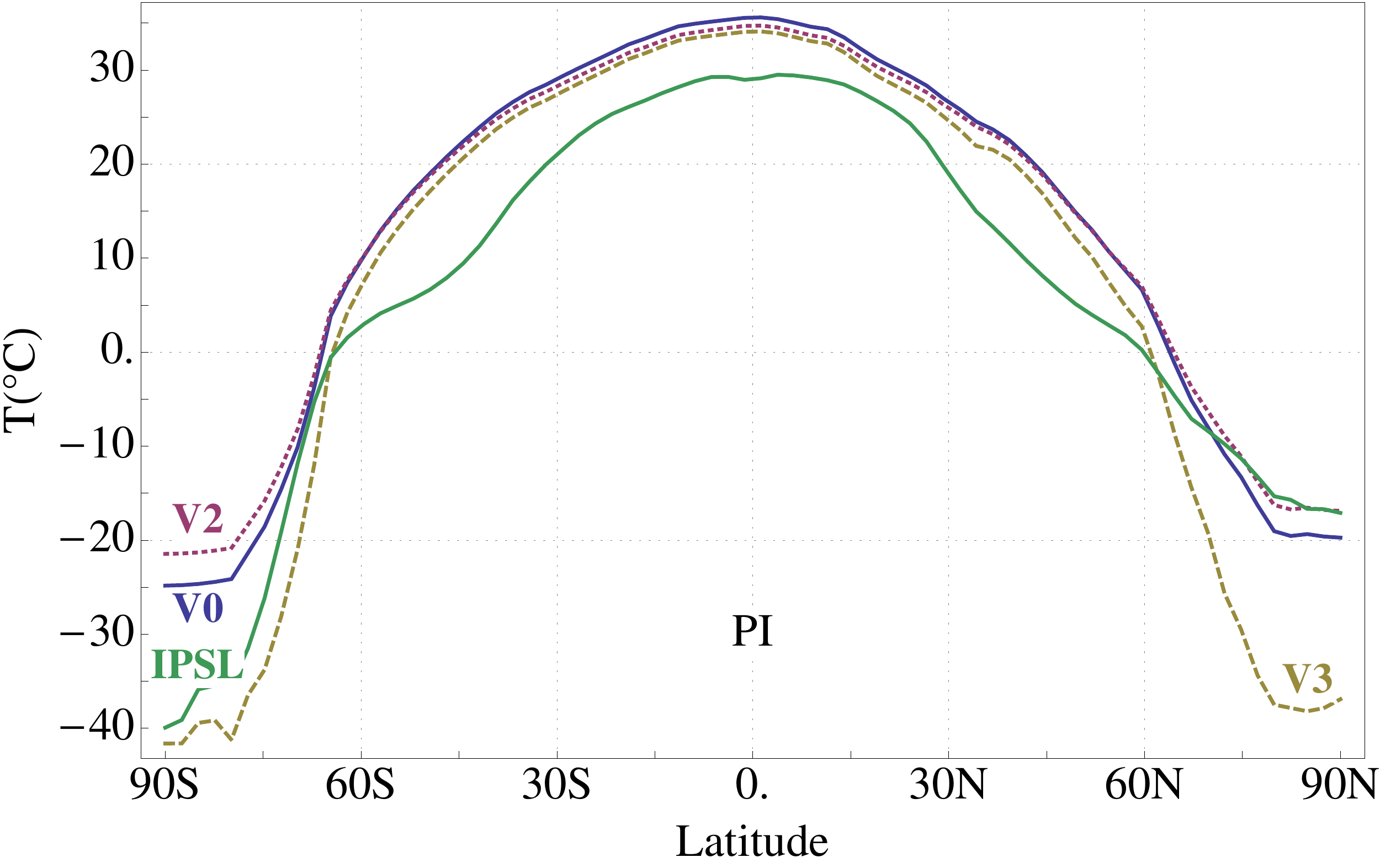}
\includegraphics[width=0.5\textwidth]{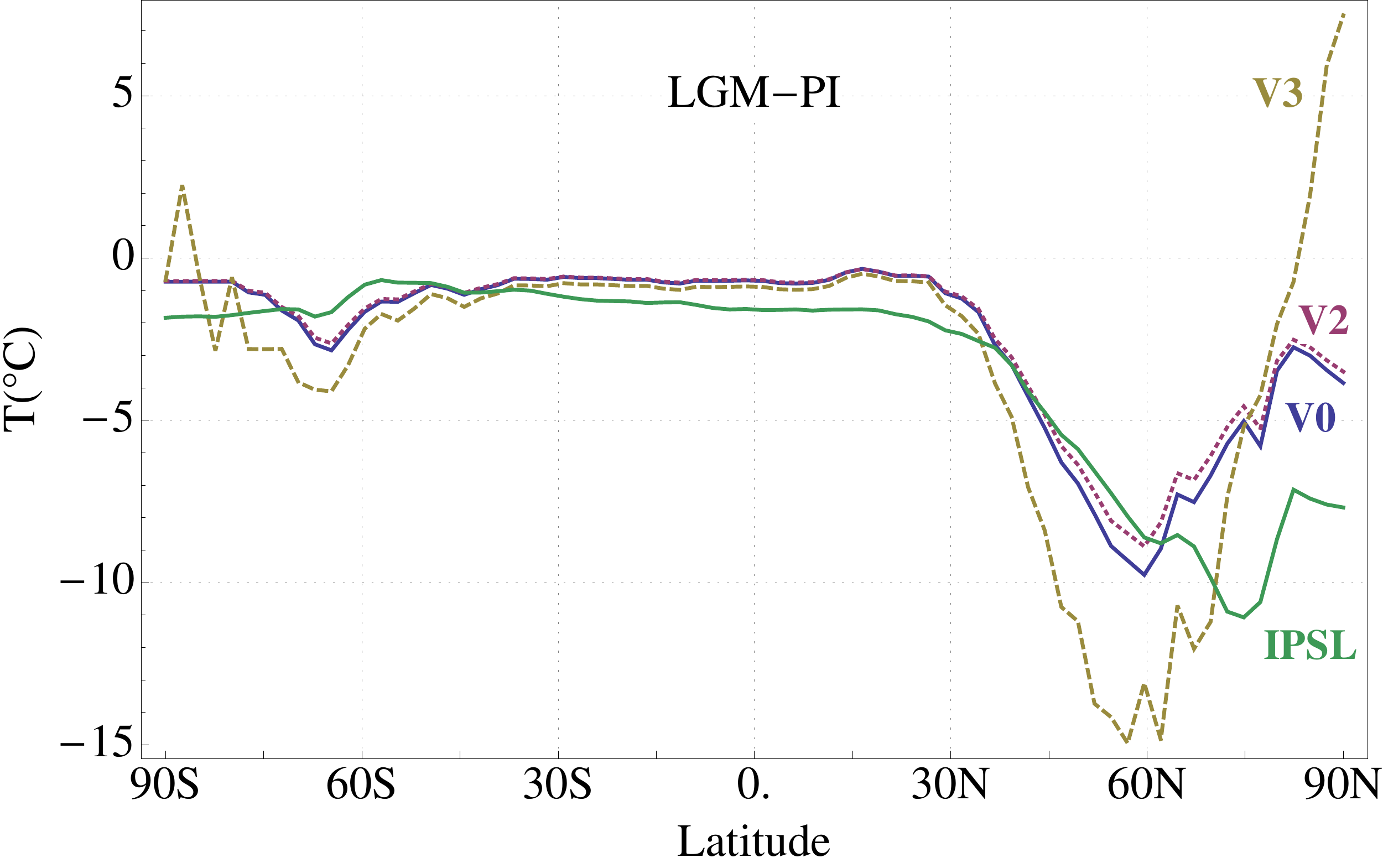}
\caption{Left: Surface temperature $T_g$ for pre-industrial conditions, for the different versions of the MEP model: version v0 (solid blue), v2 (dotted red), v3 (dashed yellow) and for the IPSL model (solid green). Right: Surface temperature difference between the Last Glacial Maximum and pre-industrial.}\label{Tsurflatfig}
\end{figure}

The latitudinal dependence of surface temperature distributions obtained from the different models\footnote{The uniform relative humidity in version 3 is chosen as the mean relative humidity in the MEP v0 case} is shown in Fig. \ref{Tsurflatfig}, for both pre-industrial and LGM conditions. When the water vapour feedback\index{Climate Feedbacks!Water-vapour Feedback} is active (version v3), the surface temperature is much lower in the polar regions than with other versions of the MEP model. For the same reason, the response to the albedo change at the LGM is also stronger (Fig. \ref{Tsurflatfig}, right). Globally, the temperature response is approximately 1 \degre C stronger than in the absence of the water vapour feedback (Table \ref{versiontable}).

\section{The importance of spatial resolution}

In the MEP procedure\index{Maximum Entropy Production}, it is traditionally argued that maximizing the entropy production constitutes a way to represent the effect of small, unresolved scales, on the large, resolved scales\index{Sub-grid Parameterizations}. In the case of meridional heat transport in (dry) planetary atmospheres, the energy is carried partly by the mean flow and partly by turbulent fluctuations. Nevertheless, even a model accounting for no dynamics at all like the MEP model shown here presents reasonable transport curves. For the sake of the comparison with the IPSL model, we started with an identical resolution for the GCM and the MEP model ($N_{lat}=72$ and $N_{lon}=96$, corresponding to a 3.7\degre$\times$ 2.5\degre grid). In the MEP model, the resolution is somewhat arbitrary as the computational cost is negligible. In the light of the interpretation of MEP as a parameterization of small-scale processes, one may naturally ask how the results of the MEP model\index{Climate Models!MEP models} depend on the resolution.

\begin{figure}
\begin{center}
\includegraphics[width=0.5\textwidth]{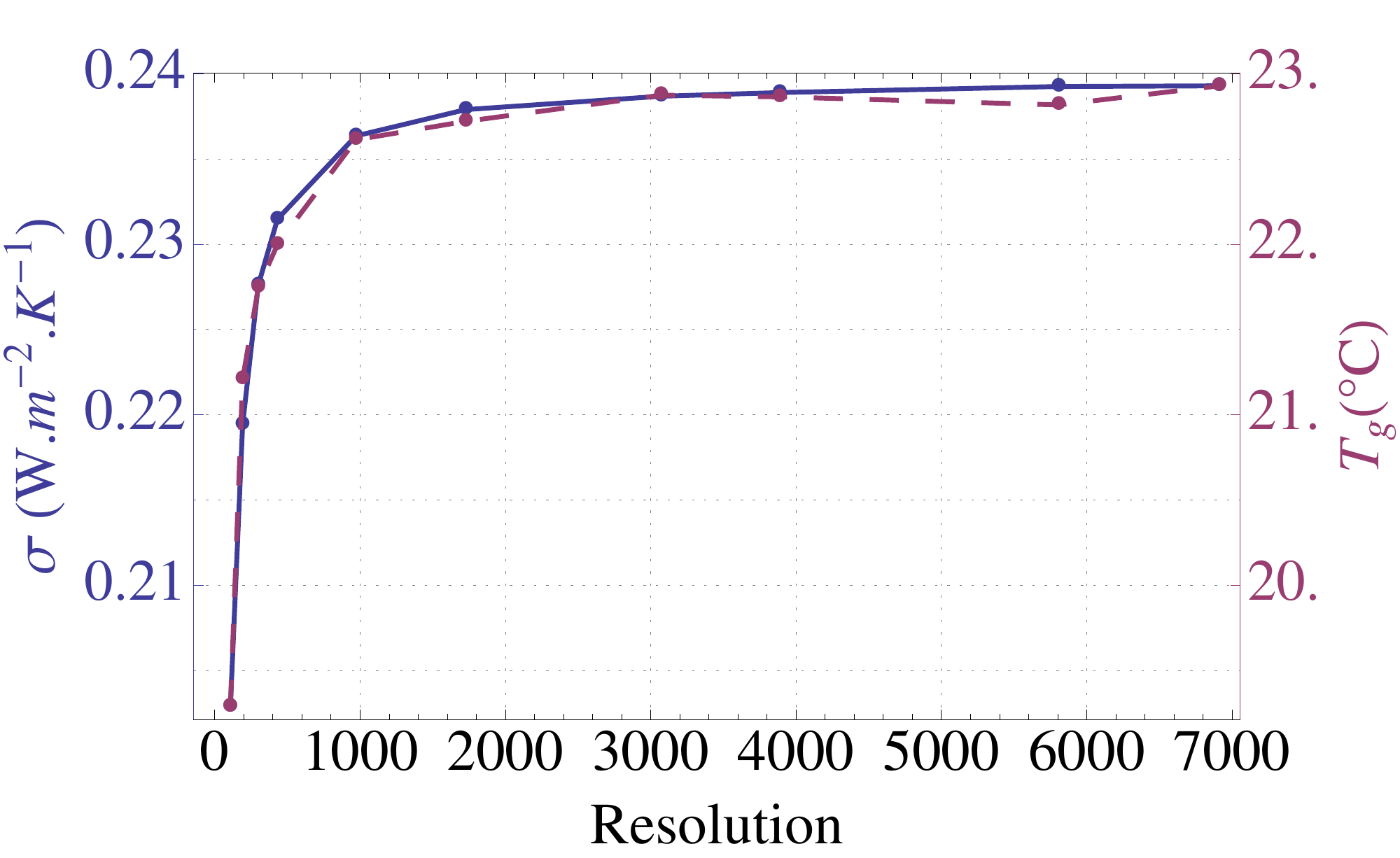}
\caption{Total material entropy production $\sigma$ (solid blue) and global mean surface temperature $\langle T_g\rangle $ (dashed red) as functions of the resolution (number of cells). The aspect ratio is maintained equal to 3/4.}\label{resfig}
\end{center}
\end{figure}

Figure \ref{resfig} shows the curves of total material entropy production\index{Entropy production} and globally averaged surface temperature obtained with the MEP v0 model with different resolutions. We keep a constant aspect ratio $N_{lat}/N_{lon}=3/4$ and vary the total number of boxes. Both curves are monotonically increasing with resolution. Although there is no explicit representation of the dynamics here, the dependence on resolution is very similar to the findings of \cite{Kleidon2003} for a GCM\index{Climate Models!General Circulation Models}. In particular, it shows that the results of the MEP model converge when the resolution increase.

\section{Future challenges for MEP climate modelling}

In this chapter, we have presented a detailed account of how the MEP conjecture\index{Maximum Entropy Production} can be applied to climate modelling. We have shown how a MEP model\index{Climate Models!MEP models} without \emph{ad-hoc} hypotheses could be built and we have compared its performances in simulating both the pre-industrial and Last Glacial Maximum climates with a coupled atmosphere-ocean GCM\index{Climate Models!General Circulation Models}. The results appear to be robust with respect to minor modifications (versions v0-v2) of the model. To go beyond these results, we argue that it is necessary to account for some feedbacks, and show how to treat them in the MEP framework. We stress the importance of the water vapour feedback (version v3) on the surface temperature\index{Climate Feedbacks!Water-vapour Feedback}. Going further would now require the ability to include a water-cycle model in our MEP model. From there one may hope to be able to represent clouds in a more robust way than in the original Paltridge model. To become a realistic climate model, the MEP model would still require important features, like a seasonal cycle, a representation of atmospheric dynamics, a more accurate description of the vertical structure, etc, but there are reasons to believe that this would not be completely out of reach. This key challenge would have to be taken up without sacrificing the original strengths of the MEP model (absence of empirical parameterizations and \emph{ad-hoc} coefficients, rapidity, conceptual simplicity). Another major point which would deserve clarification is the theoretical basis of the MEP principle\index{Maximum Entropy Production} (see \cite{Dewar2012}). In particular, it would be desirable to establish which entropy production\index{Entropy production}  should be maximized: is it always the material entropy production ? (see for instance \cite{Pascale2012}).

If this program could be achieved, the climate modelling community would acquire a valuable new tool, in addition to the existing hierarchy of models, to improve our understanding of past, present and future climates, on Earth and beyond.


\end{document}